\documentclass[aps,prl,twocolumn]{revtex4}
\usepackage{graphicx}
\usepackage{dcolumn}
\usepackage{bm}

\begin{document}
\def\be{\begin{equation}}
\def\ee{\end{equation}}
\def\bearr{\begin{eqnarray}}
\def\eearr{\end{eqnarray}}
\def\tc{$T_c~$}
\def\tcl{$T_c^{1*}~$}
\def\c2{ CuO$_2~$}
\def\ruo{ RuO$_2~$}
\def\lsco{LSCO~}
\def\bi{bI-2201~}
\def\tl{Tl-2201~}
\def\hg{Hg-1201~}
\def\sro{$Sr_2 Ru O_4$~}
\def\rc{$RuSr_2Gd Cu_2 O_8$~}
\def\mgb{$MgB_2$~}
\def\dwave{$d_{x^2-y^2}$~}
\def\blmng{$\rm{La_{2-x}Sr_{1+x}Mn_2O_7}$~}
\def\dmng{$\rm{La_{1-x}Sr_xMnO_3}$~}
\def\mng{$\rm{LaMnO_3}$~}
\def\dorb{$\rm{d_{x^2-y^2} \pm i d_{z^2}}$~}
\def\pz{$p_z$~}
\def\ppi{$p\pi$~}
\def\sqo{$S(q,\omega)$~}
\def\tperp{$t_{\perp}$~}
\def\cob{$\rm{CoO_2}$~}
\def\nxcob{$\rm{Na_x CoO_2.yH_2O}$~}
\def\ncob{$\rm{Na_{0.5} CoO_2}$~}
\def\half{$\frac{1}{2}$~}
\def\nycob{$A_xCoO_{2+\delta}$~}
\def\tj{$\rm{t-J}$~}
\def\hlf{$\frac{1}{2}$~}

\title{Latent superconductivity in doped manganites}

\author{G. Baskaran}
\affiliation{Institute of Mathematical Sciences\\
Chennai 600 113, India}

\begin{abstract}
We analyze effective Hamiltonian of ferromagnetic half metal phase 
of doped manganites and find a latent \textit{d-wave spin triplet} 
superconductivity. Spin triplet state for a d-wave pair is enabled by 
orbital degrees of freedom. This high Tc superconductivity is, however,
kept dormant by some intrinsic strong cooper pair breaking processes. 
Low T anomalies such as i) long distance superconducting proximity effects into manganites, ii) pseudogaps in tunneling, ARPES and iii) nodal quasi particles and absence of bilayer splitting in ARPES in bilayer manganite get natural and qualitative explanation. Some consequences of our orbital pairing superconductivity are 
pointed out.

\end{abstract}

\maketitle

Manganites host a variety of thermodynamic phases and exhibit rich physics
and phenomena arising from an interplay of electronic and orbital degrees of 
freedom\cite{tokura1}. The spin, charge, orbital and lattice degrees of freedom 
are tied in a complex fashion, leading to remarkable control of electron
transport by magnetic or electric fields,
in phenomena such as colossal magnetoresistance or electroresistance effects. 
These make manganites special and exciting from technological point of view. 
From basic science point of view, manganites are formidable, compared to its 
less complex cousin, high \tc copper oxides

In many transition metal oxides, orbital angular momentum of d-electrons 
are only partially quenched and some orbital degeneracy remains. Kugel and 
Khomskii\cite{khomskii1} pioneered important theoretical studies in 
this front. In doped oxides with strong electron correlations, instead of 
undergoing 
a cooperative Jahn-Teller distortion and orbital order, electrons tend 
to quantum fluctuate among degenerate 
orbitals, resulting in a state with disordered orbital occupancy. 
This state is called `orbital liquid'\cite{OrbLiq}. In the recent past, orbital fluctuation has been 
studied both theoretically and experimentally, 
leading to deeper insights into magnetic and electrical properties 
of some transition metal oxides, including violation\cite{GKviolation} of
celebrated Goodenough-Kanamori rules.

On the experimental front, fully polarized ferromagnetic manganite 
family of metals exhibit anomalous features such as large Drude width
in optical conductivity\cite{drude}, pseudo gap\cite{pgap1,pgap2,pgap3}, 
absence of quasi particle peak\cite{absentQP}, nodal quasi particles in photoemission studies\cite{nodalQP}
and anomalous superconducting proximity effects\cite{proxEff} at low temperatures. Various theoretical efforts\cite{mangTheory} have attempted to explain the rich variety of properties of manganites, occuring at small and high energy scales,
using double exchange and Jahn-Teller 
polaron mechanisms, in the ferro and paramagnetic phases. However, there is a 
need for an unified explanation of the above low T and low energy, 1 to 10 meV 
anomalies in the ferromagnetic phase.
 
The aim of the present article is to explain qualitatively low energy 
anomalies in a natural and unified fashion, for manganites and related systems. 
We suggest that in the fully spin polarized low T metallic phase there 
is a `latent high \tc superconductivity'. 
The order parameter is a d-wave spin triplet. The enhanced superconducting 
correlations arise from strong electron correlation induced pairing among 
orbital degrees of freedom between two electrons with parallel spins. 
Orbital degree of freedom offers a possibility to have 
{\em d-wave spin triplet}, 
still maintaining the overall antisymmtry of the many electron wave function.

After specializing the model Hamiltonain of manganite to fully polarized ferromagnetic phase, we isolate a major part of it. It is a large repulsive U Hubbard model (or its variant t-J model), with spin-half degree of freedom replaced by orbital (pseudospin-half) degree of freedom. This piece has global SU(2) symmetry in pseudospin space. Using an RVB 
theory analysis\cite{pwaScience,bza,krs,pVanila}, we find that this part of the model for manganite exhibits $d_{x^2-y^2}$ or $d_{y^2-z^2}$ or $d_{z^2-x^2}$
wave high \tc pseudospin (orbital) singlet superconductivity. The remaining nonsymmetric SU(2) part are strong pair breakers and destroy long range superconducting order.

We explain some experimental anomalies in a qualitative fashion using our
theory. We end with some predictions and discussions. 

Our finding also implies a generic possibility in oxides that \textit{electrons overcome single particle kinetic energy frustraion imposed by strong mutual
coulomb repulsions, by pair delocalisation in either spin or orbital channels}. It is likely that some of the known charge order correlations in manganites and other systems are finite momentum condensates of cooper pairs of the type we have suggested.  

We start with the effective Hamiltonian\cite{model1,model2} for doped 
manganite \dmng. In the Mott insulating manganite \mng
trivalent Mn$^{3+}$ has an electronic configuration 3d$^4$. In a cubic crystal field the 3d level is split into doubly degenerate e$_g$ and triply denerate
t$_{2g}$ levels. Three of the four parallel spins fill the t$_{2g}$ levels and 
the fourth occupies one of the e$_g$ doublet. Hund coupling gives a 
(maximal) spin 2 for Mn$^{3+}$ ion. In the doped metallic state Hund coupling
favors a fully polarized ferromagnetic ground state. In view of the large
Hubbard U, which discourages charge fluctuation into Mn$^{2+}$ state, 
{\em the fully polarized ferromagnetic ground state is not a simple Slater 
determinant state. It is a correlated quantum state}.  This is 
clearly suggested by the effective large U Hubbard Hamiltonian,
containing a part symmetric and rest non-symmetric, in an orbital index, 
as explained below :
\bearr
&{}&H_{{}_{\rm FM}} = H_{{}_{\rm S}}+ H_{{}_{\rm NS}}\nonumber \\
&{}&H_{{}_{\rm S}} = - t \sum_{\langle \rm{ij}\rangle \tau = \pm} 
c^{\dagger}_{i\tau \uparrow}c^{}_{j\tau \uparrow} + h.c.
+ \frac{U}{2} \sum_i n_{i+ \uparrow}n_{i- \uparrow}  \\
&{}&H_{{}_{\rm NS}} =
- t\sum_{\langle ij \rangle} e^{i\theta_{ij}} c^{\dagger}_{i+ \uparrow }c^{}_{j- \uparrow} + h.c.
\eearr

As we are in the fully polarized ferromagnetic subspace, the t$_{2g}$
core spin degree of freedom and Hund coupling do not appear 
explicitly in the Hamiltonian. 
The subscript $\tau = \pm$ refers to two complex combination of
e$_g$ orbitals: $|$d$_{x^2-y^2}\rangle \pm$ i $|$d$_{3z^2-r^2}\rangle$. 
As discussed by Feiner and Oles\cite{model2}, this complex combination gives a charge density with 
full cubic symmetry. This combination is a more natural atomic basis to 
discuss ferromagnetic manganite, around x $\approx$ 0.3, where the system 
has a cubic symmetry and is believed to be an orbital liquid, with equal
occupancy of the d$_{x^2-y^2}$ and d$_{3z^2-r^2}$ orbitals. This is inferred, 
for example, through an isotropic spin wave spectrum in neutron scattering 
results\cite{LCMOspinWave}.

Another advantage of this basis is that near neighbor hopping matrix 
elements have the same strength for both diagonal and off diagonal terms. 
However, the off diagonal term has extra phases, 
$\theta_{ij} = \pm\frac{2\pi}{3}, 0$ for hopping along a,b and c-axis,
as discussed in detail by Feiner and Oles\cite{model2}. 

For manganites band parameters estimates\cite{model1,model2} are: Hubbard U $\approx$ 
6 eV and t$\approx$ 0.4 eV. It is a large U repulsive Hubbard model, 
with a key difference of an off diagonal hopping term in the pseudo 
spin index $\tau$. That is, H$_{{}_{\rm S}}$ (eqn.1) has global SU(2) 
symmetry in pseudo spin space; H$_{{}_{\rm NS}}$ (eqn.2) does not have 
global SU(2) symmetry.

This and similar model has been studied in great detail recently\cite{model2}. 
What we offer is a new qualitative analysis and insights that has 
phenomenological support, as shown below.

The SU(2) symmetric part H$_{{}_{S}}$ is the standard large U Hubbard model 
away from half filling. RVB theory\cite{pwaScience,bza} suggests that 
antiferromagneic superexchange interaction generated by virtual fluctuations 
into doubly occupied states lead to spin singlet correlations and superconductivity. In our case it gives a superconducting ground state with 
cooper pairs made of orbital pseudospin singlets and physical spin 
triplets  $|\uparrow\uparrow\rangle$. The corresponding electron
pair is created by an operator,
\be
b^{\dagger}_{ij} \equiv 
\frac{1}{\sqrt 2} (c^{\dagger}_{i+\uparrow}c^{\dagger}_{j-\uparrow}- c^{\dagger}_{i-\uparrow}c^{\dagger}_{j+\uparrow}) 
\ee
A very successful variational ground state for the SU(2) symmetric 
part of the Hamiltonian is the d-wave superconducting state, containing 
a Gutzwiller-Jastrow factor which discourages double occupancy: in our
present case it is a spin triplet, pseudo spin singlet d-wave 
superconducting state:
\be
|G\rangle = \prod_i ( 1 - g n_{i+ \uparrow} n_{i- \uparrow}) 
\prod_k (u_k + v_k c^{\dagger}_{-k+\uparrow} c^{\dagger}_{k-\uparrow})|0\rangle
\ee
The cooper pair function $\frac{v_k}{u_k}$ has d$_{x^2-y^2}$ type symmetry.
The projection factor g = 1 - O($\frac{t}{U}$). 

As we have a 3 dimensional system, there are three degenerate states of
the d-wave superconductor order parameter: d$_{x^2-y^2}$, d$_{y^2-z^2}$ and d$_{z^2-x^2}$.
In the latent superconducting phase, all three pairing
will appear with equal probability, as fluctuating domains
of superconducting pair correlations. For the case of bilayer manganite,
\blmng, the intrinsic 2D character will pickup only the d$_{x^2-y^2}$ 
pairing correlations. The magnitude of the superconducting gap depends
on doping. For manganites in the good metallic ferromagneic state x 
$\approx$ 0.3, we are in the overdoped regime, as compare to layered
cuprates.

This seemingly robust long range superconducting order is 
not stable against pair breaking processes represented by the off diagonal 
hopping term H$_{{}_{\rm NS}}$. This term flips the pseudo spin and is analogous to a `spin-orbit' (pseudospin-orbit) coupling, but strong. A pseudo spin singlet electron pair has a finite probability of becoming pseudo spin triplet 
spontaneously.  In an analysis, similar to pair breaking theory in BCS theory, we find that superconducting Tc gets suppressed to zero. 

It is somewhat straight forward to see that the strong pair breaking 
destroys d-wave superconductivity. However, it is difficult to quantify the surviving pairing correlations. The problem is hard and has some similarity  
to issues of pairing correlations in the spin gap phase in cuprates.

Jahn-Teller coupling leads to cooper pair breaking, as outlined below.
Coupling of charge of an e$_g$ electron to c-axis (or a or b axis) distortion of oxygen octahedra lifts two fold e$_{g}$ degeneracy, as represented by the electron-lattice interaction Hamiltonian:
\be
H_{{}_{\rm JT}} = g_{{}_{\rm JT}}\sum_i (c^{\dagger}_{i1\uparrow}c^{}_{i1\uparrow}
-c^{\dagger}_{i2 \uparrow}c^{}_{i2 \uparrow})~ {\hat{q_i}}
\ee
Here the subscript 1,2 refer to the d$_{x^2-y^2}$ and d$_{3z^2-r^2}$ orbitals;
$\hat{q_i}$
is the c-axis distortion variable and Jahn-Teller coupling constant $ g_{{}_{\rm JT}}
\sim $ 10 eV/Au. After transforming to the complex d $\pm$ id basis, it becomes an off diagonal scattering term, that flips the pseudo spin:

\be
H_{\rm JT} =  g_{{}_{\rm JT}}\sum_{i}(c^{\dagger}_{i+ \uparrow}c^{}_{i- \uparrow}
+c^{\dagger}_{i- \uparrow}c^{}_{i+ \uparrow}) ~{\hat{q_i}}
\ee

Thus the major Jahn-Teller coupling does not conserve the psuedo spin and
hence acts like a pair breaker for the pseudo spin singlet cooper pairs.
However, coupling to some specific inter octahedral distortions may favour 
local bond singlets of pseudospins.

As we increase T, thermal spin waves (goldstone modes) are produced. 
The more the occupancy of the spin wave modes, the more we go out of the ferromagnetic Hilbert space. The pseudo spin singlet stabilizing process 
is compatible with Hund coupling; that is, once neighboring spins are antiparallel, Hund coupling discourages charge hopping and hence 
pseudo spin `superexchange'. Thus electron spin reversals reduce the 
pseudospin singlet or superconducting correlation. 

Having suggested a possibility of enhanced superconducting correlation
in the ferromagnetic metallic doped manganites, we will discuss known 
anomalies in the present light.  They seem to fall in place, at a 
qualitative level.

When a normal metal is in contact with a singlet or triplet superconductor, proximity effect\cite{deGennes} is induced over a length scale given by 
$\xi \sim \sqrt{\frac{\hbar v_F \ell_F}{\pi kT}}$ (dirty limit) or 
($\frac{\hbar v_F}{6\pi kT}$) (clean limit). Here $\ell_F$ is the electron
mean free path at the fermi surface. The diverging proximity effect, 
reflects divergent singlet and triplet pair susceptibility of a free fermi gas 
as T $\rightarrow$ 0. If the metal is replaced by a half metallic ferromagnet (LCMO) and the superconductor by a singlet superconductor (YBCO), singlet pair response in the metal is no longer divergent and the proximity length scale 
becomes $\xi \sim \sqrt{\frac{\hbar v_{F} \ell_F}{\pi E_{\rm ex}}} < $ 1 nm, 
a non divergent quantity at low temperatures.
This is because a low energy spin singlet cooper pair is incompatible with
a ferromagnetic half metal, where the exchange splitting E$_{\rm ex}$
makes down spin energy very high. However, in some hetrostructurers, supercurrent between singlet superconductors is seen to flow through spin polarized ferromagnetic manganitic layers\cite{proxEff} of large thickness of 
20 to 100 nm. This is an anomaly, unexpected in a dirty metal like manganite.

It has been suggested\cite{proxTheory} that at the interface there is a conversion of spin singlet to triplet cooper pairs through spin flip processes as well as
spin rotation accompanying reflection of electron\cite{sauls} by spatially varying exchange field across the interface. Spin triplet is compatible with a fully polarized ferromagnetic state; in principle this induced triplet pair correlation in the vicinity of the interface is capable of extending
into ferromagnetic half metal, by standard proximity effect.
However, ferromagnetic manganite is not a standard fermi liquid metal\cite{TVR1}; electrons undergo strong scattering through
electron-electron and electron-phonon scattering. This is evident in optical conductivity, for example, where one sees an anomalously large Drude 
width\cite{drude}. We cannot use standard proximity theory of normal metals. We suggest that an enhanced orbital d-wave pairing and availability of the new orbital channel might cause the anomalous proximity effect, as argued below.

An electron quasiparticle from normal metal undergoes Andreev reflection,
at the superconductor-normal metal interface and becomes a hole with opposite spin:

$|{\bf k}\uparrow\rangle_{\rm e} \rightarrow 
|-{\bf k}\downarrow \rangle_{\rm h}$. This creates an Andreev bound state and 
thereby an anomalous pair amplitude gets built inside the normal metal between
two superconductors.
Andreev bound state carries the Josephson current and creates proximity effect.
We suggest the following Andreev reflection process for our superconductor-
manganite metal interface: 
\be|{\bf k}\uparrow + \rangle_{\rm e} \rightarrow 
|-{\bf k} \uparrow - \rangle_{\rm h}
\ee
That is, an electron with upspin ($\uparrow$) and pseudo spin $\tau = + $
gets reflected at the boundary to become a hole with upspin ($\uparrow$)
and pseudo spin $\tau = - $. In a spin singlet superconductor
electron reverse the spin in an Andreev reflection. However,
as mentioned before, appearance of a finite exchange field on one side of the interface removes the global spin SU(2) symmetry. That is why the process represented by equation (5) will occur with a finite probability that is related to spin rotation by exchange field as well as spin flip scattering by magnetic impurities. In a similar fashion there are processes that causes pseudospin
flip, required by orbital singlet pairing in manganite. 

Qualitatively, making use of the existing latent superconductivity in 
manganite, the above Andreev reflection builds a stronger spin triplet, orbital 
singlet anomalous pair amplitude and longer proximity effect, than expected
in a dirty metallic state of manganite.

Different pseudogaps have been seen in STM, tunneling and ARPES
\cite{pgap1,pgap2,pgap3,absentQP}, over two energy scales, one around 
200 meV and the other\cite{pgap2} around 15 meV. The origin of the very
large gaps in tunnelling and STM is not very clear. We suggest that the smaller pseudogap seen in Pb-manganite tunnelling\cite{pgap3} corresponds to our orbital 
pairing and a gap of 15 meV is consistent with t, U and a large doping density 
for manganites. 

Following earlier studies\cite{absentQP}, a recent ARPES\cite{nodalQP} 
shows the following striking phenomena in bilayer \blmng
at very low energy scales, of the order of 10 meV or less: i) carriers are 
strongly scattered at the fermi surface, electron spectral weight is strongly
suppressed, resembling the normal state pseudogap phase of doped cuprates, 
ii) a quasi particle type structure emerge, surprisingly in the nodal (0.0)-($\pi,\pi$) direction in k-space, once again resembling the pseudo gap 
phase of layered cuprates and iii) absence of bilayer splitting.

At a qualitative level all the above three anomalies become less anomalous
and somewhat natural, in the light of our latent superconductivity with d$_{x^2-y^2}$ symmetry in bilayer manganite. In the RVB theory, nodal quasi particles reflect persistence of d-wave superconducting correlations, nodal
Bogoliubov like quasi particle excitations and  
spin singlet pairing into normal state. The low temperature part of the pseudo
gap region is where anomalous Nernst signal and local superconductivity,
sufficient to support long lived vortices are seen. Above this region, 
in the high temperature pseudo gap region we have spinon pairing. It will be
important to get further information experimentally on the pseudo gap phase
in manganite and its bilayer version, looking for some key similarities with
cuprates.

Absence of bilayer splitting is related, in our proposal, to 
`c-axis confinement', well known in cuprates\cite{pwaBook}  
However, in \blmng bare c-axis 
matrix element (within a bilayer) has same magnitude as the ab-plane one, 
unlike bilayer cuprates, where it is small by an order of magnitude. This makes 
the confinement in bilayer manganites even more striking and special.

The strong coupling to phonon and dynamical Jahn-Teller phenomena must be 
playing important role in manganites even in the ferromagnetic metallic 
phase, but at higher energy scales $\sim$ 100 meV. What we have suggested
is that very low energy features such as anomalous proximity effects and 
presence of nodal quasiparticles are governed mostly by strong correlation 
physics and superconducting pairing correlations advocated in the present 
paper. 

Some interesting consequences of our proposals are: i) orbital singlet 
correlation with d-wave symmetry will give rise to a pseudospin-1, 
`orbiton' excitation\cite{orbiton} (similar to 41 meV resonance in cuprates) 
at wave vectors $(\pi,\pi,\pi)$ in manganites and at $(\pi,\pi)$ in bilayer 
manganite. 
ii) Strain and pressure control of orbital fluctuation: epitaxial 
strain, for example, is like to have strong effect on the anomalous 
proximity effect
and iii) Charge 2e pairing in the latent superconducting state is likely to 
leave its signature in noise experiments in nano manganite systems. 

The pseudospin symmetry breaking term H$_{{}_{\rm NS}}$ has no independent 
parameter. Because of this we have no separate experimental control over the 
pair breaking processes. 

In conclusion, using a body of already existing theoretical and experimental
insights in cuprates and manganites, we have made a novel proposal.
It will be interesting to develop our theoretical proposal further and 
study experimental consequences. The physics we have advocated for 3D 
manganites is likely to have some key common features with CrO$_2$, another 
remarkable half metal and perhaps other oxides.

I thank, G. Khaliullin, N. Kumar, G.V. Pai and T.V. Rakmakrishnan for 
discussion on orbital dynamics in oxides over years, D.I. Khomskii for an early discussion on importance of $|$d$_{x^2-y^2}\rangle \pm$ i $|$d$_{3z^2-r^2}\rangle$
atomic orbitals for manganites, A. Roychoudhuri for reference 8 and a discussion
and Hyun-Tak Kim for an early correspondence\cite{HTkim} on p-wave superconductivity in doped manganite.


\begin{references}

\bibitem{tokura1} Y. Tokura, Rep. Prog. Phys.. {\bf 69} 797 (2006);
A.M. Oles, Phys. Stat. sol., {\bf B 236} 288 (2003);
E. Dagotto, T.~Hotta, and A. Moreo, Phys. Rep. {\bf 344} 1 (2001);
Y. Tokura and N. Nagaosa, Science, {\bf 288} 462 (2000);
Physics Of Transition Metal Oxides, Edited by S. Maekawa and G. Khaliullin
(Springer Verlag)2004. 

\bibitem{khomskii1} K. I. Kugel and D.I. Khomskii, Sov. Phys. JETP.,
{\bf 37} 725 (1973) and Sov. Phys. Usp., {\bf 25} 231 (1982)

\bibitem{OrbLiq} G. Khaliullin and S. Maekawa, Phys. Rev. Lett. {\bf 85} 3950 (2000);S. Ishihara, M. Yamanaka, and N. Nagaosa, Phys. Rev., {\bf B56} 686 (1997);
J. Zaanen and A.M. Oles, Phys. Rev., {\bf B 48} 7197 (1993); 

\bibitem{GKviolation}A. M. Oles et al., Phys. Rev. Lett., {\bf 96} 147205 (2006)

\bibitem{drude}
M. Quijada et al., \prb {\bf 58}, 16093 (1998); E Saitoh et al., J. Phys. Soc. Jap., {\bf 69} 3614 (2000)

\bibitem{pgap1}
J. Y. T. Wei, N.-C. Yeh, and R. P. Vasquez
\prl {\bf 79} 5150(1997)

\bibitem{pgap2}H. M. Ronnow et al., Nature {\bf 440} 1025 (2006)

\bibitem{pgap3}J. Mitra et al., \prb {\bf 68} 134428 (2003)

\bibitem{absentQP} Dessau, D. S. et al. \prl {\bf 81} 192 (1998)

\bibitem{nodalQP} N. Mannella et al., Nature, {\bf 438}474 (2005)
\bibitem{proxEff}
V. Pena et al.,\prb {\bf 69} 224502 (2004);
Z. Sefruiyu et al., \prb {\bf 67} 214511 (2003)

\bibitem{mangTheory}
C. Zener, Phys. Rev., {\bf 82} 403 (1951);
P.W. Anderson and H. Hasegawa, Phys. Rev., {\bf 100} 675 (1955)
N. Furukawa, J. Phys. Soc. Jpn. 64, 2734 (1995); 
A. J. Millis, P. B. Littlewood, and B. I. Shraiman, 
\prl {\bf 74} 5144 (1995); C.M. Varma, \prb {\bf 54} 7328 (1996);
T. V. Ramakrishnan et al., \prl {\bf 92} 157203 (2004);
A. S. Mishchenko and N. Nagaosa, \prl {\bf 93} 036402 (2004)

\bibitem{pwaScience} P.W. Anderson, Science {\bf 235} 1196 (1987).
\bibitem{bza} G. Baskaran, Z. Zou and P.W. Anderson,  Sol.
State Commm. {\bf 63} 973 (1987); G. Baskaran and P.W. Anderson, 
Phys. Rev. B {\bf 37} 580 (1988)
\bibitem{krs} S. Kivelson, J. Sethna and D. Rokshar, Phys. Rev. B
{\bf 35} 8865 (1987); G. Kotliar, \prb {\bf 37} 3664   (1988)
\bibitem{pVanila} P. W. Anderson et al., J. Phys.: Condens. Matter 
{\bf 16} R755 (2004)

\bibitem{41meV}	H.F. Fong et al., Phys. Rev. Lett. 75, 316 (1995)

\bibitem{model1} S. Ishihara, J. Inoue, and S. Maekawa, \prb {\bf 55} 8280 (1997); R. Shiina, T. Nishitani, and H. Shiba, J. Phys. Soc. Jpn. {\bf 66} 3159 (1997)
\bibitem{model2} L. F. Feiner et al., \prb {\bf 71}, 144422(2005)
\bibitem{LCMOspinWave}T.G. Perring et al., \prl {\bf 77} 711 (1996)
\bibitem{deGennes} Theory of Superconductivity in metals and alloys,
de Gennes, (W.A. Benjamin, NY) 1966
\bibitem{proxTheory}
M. Eschrig et al., \prl {\bf 90} 137003 (2003)
F.S. Bergeret et al., Rev. Mod. Phys., {\bf 77} 1321 (2005)
\bibitem{sauls} T. Tokuyasu, J.A. Sauls and D. rainer, \prb {\bf38} 8823 (1988)
\bibitem{TVR1} T.V. Ramakrishnan, Philos. Trans. R. Soc. London, Ser. A 356, 41 (1998)
\bibitem{pwaBook} The theory of high Tc superconductivity,
P.W. Anderson, (Princeton Univ. Press, Princeton) 1997
\bibitem{orbiton} E. Saitoh et al., Nature {\bf 410} 180 (2001)
\bibitem{HTkim} H.T. Kim informed me (private communication, 1998) of his
preliminary experimental evidence for p-wave low-T superconductivity in doped manganites.
\end{references}
\end{document}